\documentclass[lettersize,journal]{IEEEtran}
\usepackage{amsmath,amsfonts}
\usepackage{algorithmic}
\usepackage{array}
\usepackage[caption=false,font=normalsize,labelfont=sf,textfont=sf]{subfig}
\usepackage{textcomp}
\usepackage{stfloats}
\usepackage{url}
\usepackage{verbatim}
\usepackage{graphicx}
\def\BibTeX{{\rm B\kern-.05em{\sc i\kern-.025em b}\kern-.08em
    T\kern-.1667em\lower.7ex\hbox{E}\kern-.125emX}}
\usepackage{balance}
\begin{document}
\title{Micro-expression recognition \\based on depth map to point cloud}
\author{Ren Zhang, Jianqin Yin, Chao Qi, Zehao Wang, Zhicheng Zhang, Yonghao Dang
\thanks{The authors are with the School of Artificial Intelligence, Beijing University of Posts and Telecommunications, Beijing 100876, China (e-mail: zhangren@bupt.edu.cn; jqyin@bupt.edu.cn; qichao199@163.com; zczhang@bupt.edu.cn; 2199586379@qq.com; dyh2018@bupt.edu.cn;).} }

\maketitle

\begin{abstract}
Micro-expressions are nonverbal facial expressions that reveal the covert emotions of individuals, making the micro-expression recognition task receive widespread attention.  However, the micro-expression recognition task is challenging due to the subtle facial motion and brevity in duration. Many 2D image-based methods have been developed in recent years to recognize MEs effectively, but, these approaches are restricted by facial texture information and are susceptible to environmental factors, such as lighting. Conversely, depth information can effectively represent motion information related to facial structure changes and is not affected by lighting. Motion information derived from facial structures can describe motion features that pixel textures cannot delineate. We proposed a network for micro-expression recognition based on facial depth information, and our experiments have demonstrated the crucial role of depth maps in the micro-expression recognition task. Initially, we transform the depth map into a point cloud and obtain the motion information for each point by aligning the initiating frame with the apex frame and performing a differential operation. Subsequently, we adjusted all point cloud motion feature input dimensions and used them as inputs for multiple point cloud networks to assess the efficacy of this representation. PointNet++ was chosen as the ultimate outcome for micro-expression recognition due to its superior performance. Our experiments show that our proposed method significantly outperforms the existing deep learning methods, including the baseline, on the $CAS(ME)^3$ dataset, which includes depth information.
\end{abstract}

\begin{IEEEkeywords}
Micro-expression, facial expressions, depth information
\end{IEEEkeywords}

\section{Introduction}
\begin{figure}[h]
  \centering
  \includegraphics[width=\linewidth]{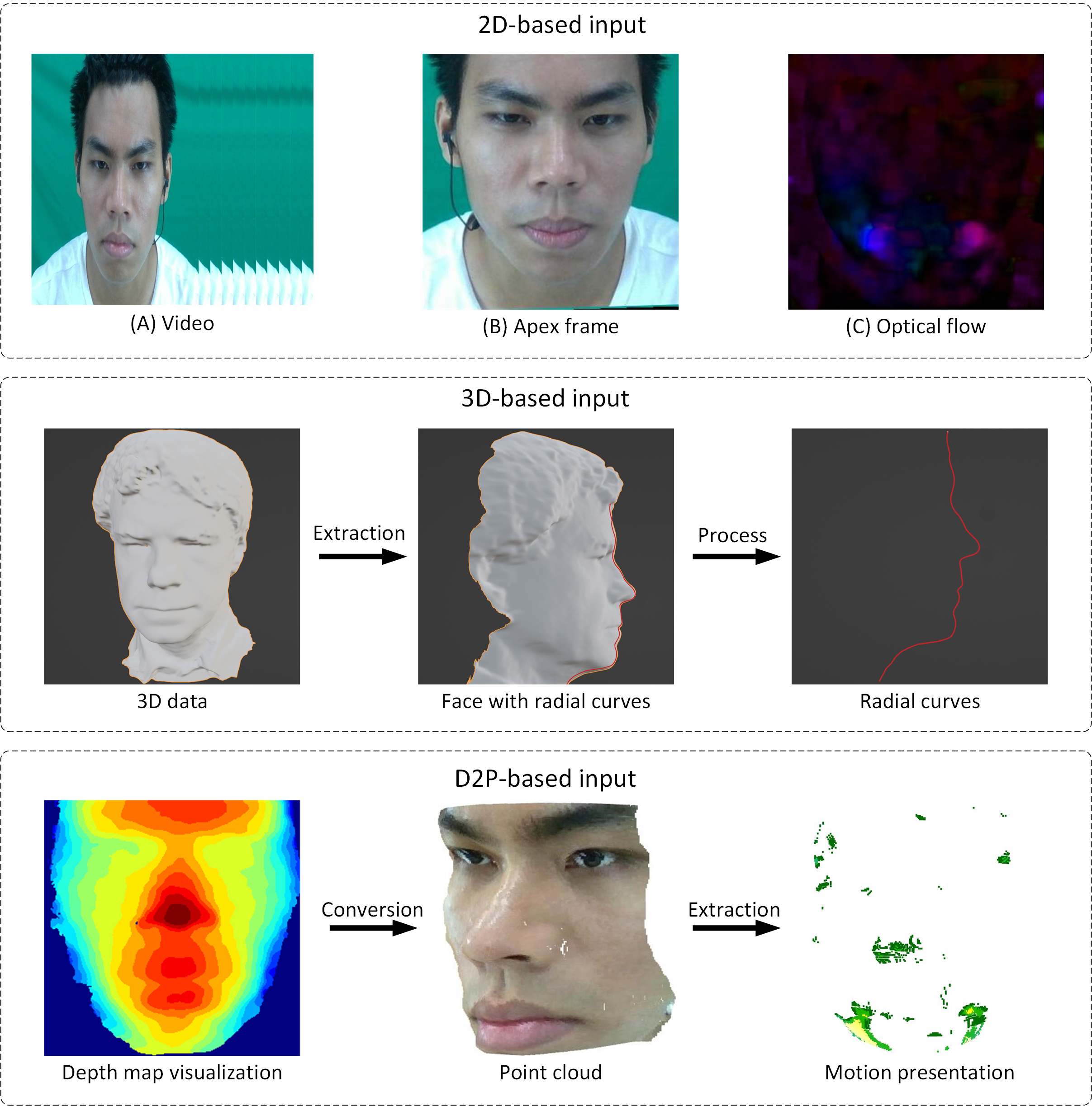}
  \caption{Input data types. Row 1 (A), (B), and (C) correspond to the input data for traditional MER methods: video, apex frame, and optical flow. Row 2 signifies the facial 3D data commonly used for macro-expression recognition that detects changes in facial surface curves to recognize emotion categories. Row 3 indicates that facial motion features are preserved by converting depth maps to facial point clouds for the MER.}
  \label{fig:intro}
\end{figure}

\begin{figure*}[h]
  \centering
  \includegraphics[width=\linewidth]{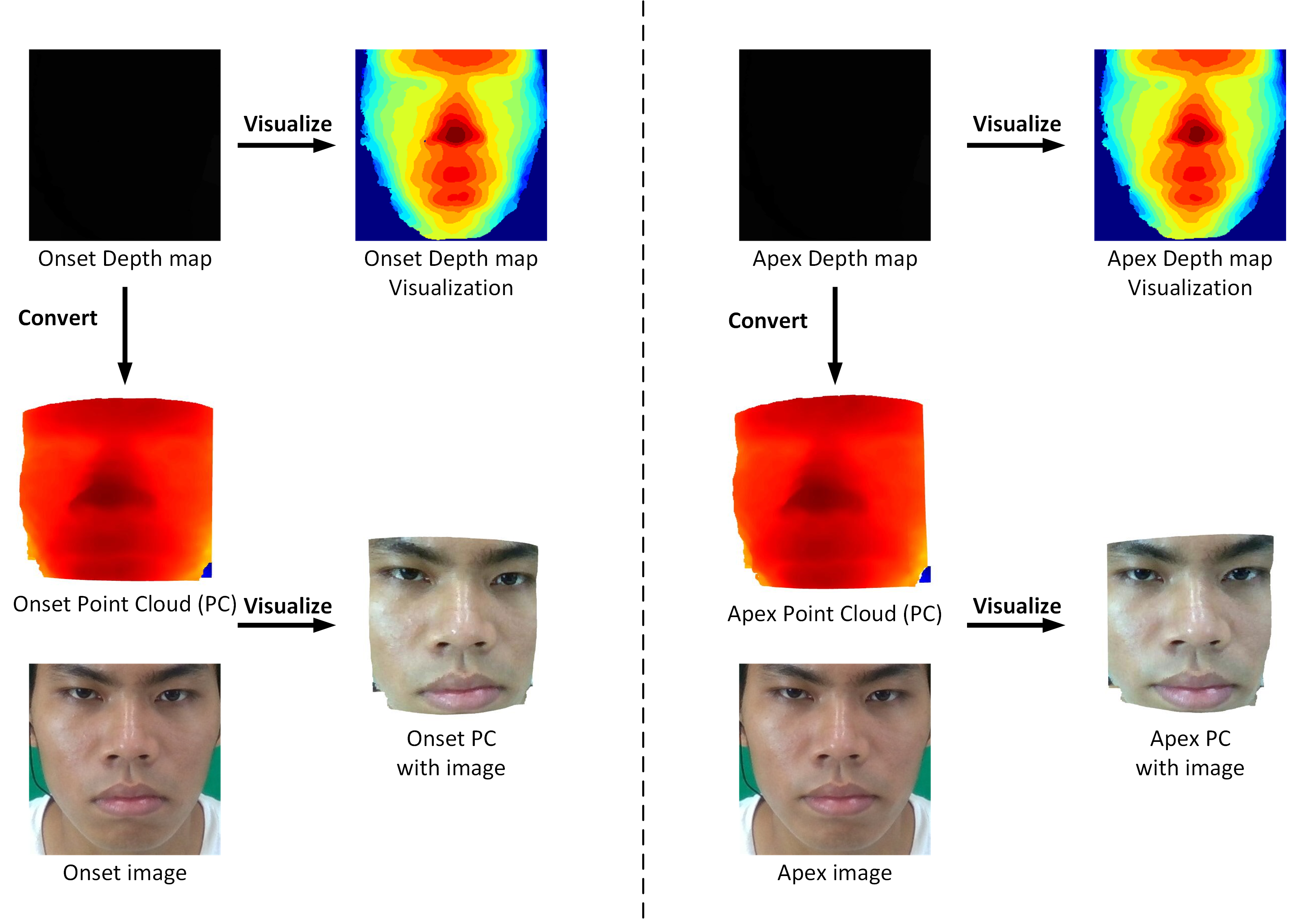}
  \caption{The point cloud aligns with the image. The depth map aligns perfectly with the image; converting the depth map to a point cloud aligns the image with the point cloud,  allowing us to crop the point cloud using the image and obtain a preliminary facial structure.}
  \label{fig:0}
\end{figure*}

Facial expressions are a key source of information for judging emotions, understanding social interactions, and inferring mental states\cite{ekman2003darwin}. There are two types of expressions, micro- and macro-expressions, depending on the duration of the expressions. Usually, the macro-expression(MaE) appears more frequently and last longer than the micro-expression(ME) and is therefore easy to spot \cite{ahmed2021deception}. On the other hand, ME is brief, subtle, localized, and less likely to be perceived\cite{li2022cas}. It stems from a failed attempt to hide or suppress emotions\cite{frank2015microexpressions}. Generally, MEs are difficult to detect and recognize, with only 40\% accuracy for untrained individuals\cite{weiss2011ekman}. ME understanding is critical in many applications, mainly lie detection, which is crucial in criminal analysis.\\
Tremendous progress has been made in image-based micro-expression recognition(MER) methods which typically have three input types: continuous video frames\cite{pfister2011recognising} (Fig. \ref{fig:intro}.(A)), apex frames\cite{wang2020micro} (Fig. \ref{fig:intro}.(B)), and optical flow
features\cite{liu2019neural} (Fig. \ref{fig:intro}.(C)). Using video sequences and apex frames are apparent-based methods focusing on pixel and texture features, and using optical flow are geometric-based methods focusing on the movement of feature points or regions.\\
However, methods using the three inputs above are limited by the texture and quality of the image; and suffer from illumination and pose variations that often occur in real life. With more modal data captured, facial structure motion information is a promising alternative for achieving robust MER. Indeed, facial structure motion feature (including depth and 3D data) directly reflects the facial surface deformation caused by muscle movements \cite{7457243}, which is highly related to facial expressions. By utilizing changes in the facial surface curve, the position and trend of facial movements can be obtained (Fig. \ref{fig:intro} line 2). Moreover, they are immune to changes in lighting and viewpoint and can overcome some limitations of image representations, such as sensitivity to lighting conditions, pose, and makeup. Specifically, some muscular facial action units are difficult to distinguish by images, and their 3D representation can overcome this limitation, which has been widely used in many macro expression recognition methods\cite{SANDBACH2012683}.\\
In the context of the MER task, variations in traditional 3D facial surface curves struggle to differentiate subtle facial movements effectively. Compared to 3D data, depth maps can more accurately capture the movement of individual facial points. Experimental evidence demonstrates that incorporating depth information in the context of the MER method is effective in the MER task \cite{li2022cas}. Using depth maps as input introduces challenges, including excessive redundancy and ambiguous structure. However, with suitable camera parameters, the depth information of the face allows for a non-destructive restoration of the facial structure. It offers more precise and comprehensive positional information between points. Furthermore, by removing stationary points, complete retention of the facial motion information minimizes interference from redundant points. (Fig. \ref{fig:intro} line 3). Moreover, the facial structure converted from depth information has a natural advantage. Each point in the point cloud aligns with corresponding pixels in the 2D image, and each point in two frames of the same video sequence is consistently located, as shown in Fig.  \ref{fig:0}. Visualization results demonstrate that the point cloud provides precise spatial information and facilitates the alignment of distinct facial regions across frames.\\
To better utilize depth information for representing facial structure and acquiring facial motion information, we first set the camera focal length to convert the onset and apex frame depth maps into point clouds containing accurate facial structure information. To obtain motion information for each pixel on the face, we calculate the regional difference of the corresponding point clouds in the onset frame and apex frame. Additionally, to describe motion information related to micro-expressions, we use coordinate transformation to determine the magnitude and direction of motion. Subsequently, we normalize and sort the motion representation of each point, retaining sufficient motion information to capture the motion characteristics of the video segment. Finally, by preserving the complete facial motion information, we have attempted various point cloud networks to model the relationship between motion features and emotional categories. And we verify that our proposed point cloud characterization can be achieved across multiple models beyond the current state-of-the-art methods and baselines. The specific innovation points are as follows:\\
1. We propose a novel paradigm for utilizing depth information in MER, effectively representing facial motion features in a simplified manner. By converting depth information into point clouds and calculating motion features, we can filter and exclude non-moving point clouds from facial data, effectively characterizing facial motion features with minimal computational cost.\\
2. Our proposed method of representing micro-expressions was validated on multiple point cloud models to evaluate its effectiveness. We conducted modeling of the relationship between local and global motion features of micro-expression point clouds and expression categories by utilizing multiple point cloud models.The obtained recognition results were significantly better than the current state-of-the-art micro-expression methods, further providing evidence for the effectiveness of the micro-expression point cloud representation method.\\
3. We proposed the novel depth-based MER method that exhibits superior performance compared to contemporary methods on the $CAS(ME)^3$ dataset, achieving state-of-the-art results across multiple categories. The experiments validate the substantial contribution of our proposed motion representation method in accurately capturing fine-grained motion positions.
\section{Related Work}
Facial expression recognition(FER) has been widely appreciated as an important method for identifying human emotional expressions. In the early days, the focus was more on easy-to-find macro expressions. MEs were first discovered in 1966 \cite{haggard1966micromomentary}. Three years later, Ekman et al. used MEs to analyze a video interview of a patient who had attempted suicide \cite{ekman1969nonverbal}. With the proposed Local Binary Patterns from Three Orthogonal Planes (LBP-TOP) methods\cite{pfister2011recognising}, MER by computer has become possible. In the past decade, many deep learning methods already exist to recognize MEs with good recognition results\cite{zr2021MER}. However, due to the lack of samples, there are few methods to recognize MEs using depth or 3D information. On the contrary, more methods for 3D FER are available\cite{alexandre2020systematic}.
\subsection{MER with image}
One of the challenges of using deep learning to recognize MEs is that the number of samples is small, and the network is exceptionally prone to overfitting. As the number of ME datasets grows, more and more work is starting to emerge. The MEGC competition combines the SMIC \cite{li2013spontaneous}, SAMM \cite{davison2016samm}, and CASME II \cite{yan2014casme}datasets to construct a larger ME sample and employs Unweighted F1-score ($UF1$) and Unweighted Average Recall ($UAR$) as evaluation metrics. The competition uses LBP-TOP as a baseline, treats video as a 3D structure with X, Y, and T coordinates, and extracts LBP features on three types of planes (XY, XT, YT) separately for MER.LBP-TOP is a manual feature method that does not depend on sample size and can be competent for early MER tasks. Several ways give the same answer for solving the problem of small ME samples. Transfer learning reduces the number of training sessions and prevents overfitting by transferring the knowledge learned on a large sample to a small sample. The Micro-Attention method \cite{wang2020micro} takes the apex frames of ME videos as input, designs a unique network structure ResNet10 to reduce the occurrence of overfitting, and designs micro-attention so that the network focuses on regions of interest of faces covering different actions. After pre-training on several macro-expression datasets, and fine-tuning the ME dataset achieves good recognition results, migration learning in this method achieved the champion of MEGC2018 \cite{Transfer2018}.\\
Conversely, the EMR method \cite{liu2019neural} focuses on the difference in the movement amplitude between micro- and macro-expressions. The EMR method proposes expression magnification and reduction, i.e., magnifying MEs and reducing the magnitude of ME movements. It introduces an adversarial-based domain adaptation technique reducing the difference between the two domains. With this idea, the method won the MEGC 2019.
\begin{figure}[h]
  \centering
  \includegraphics[width=\linewidth]{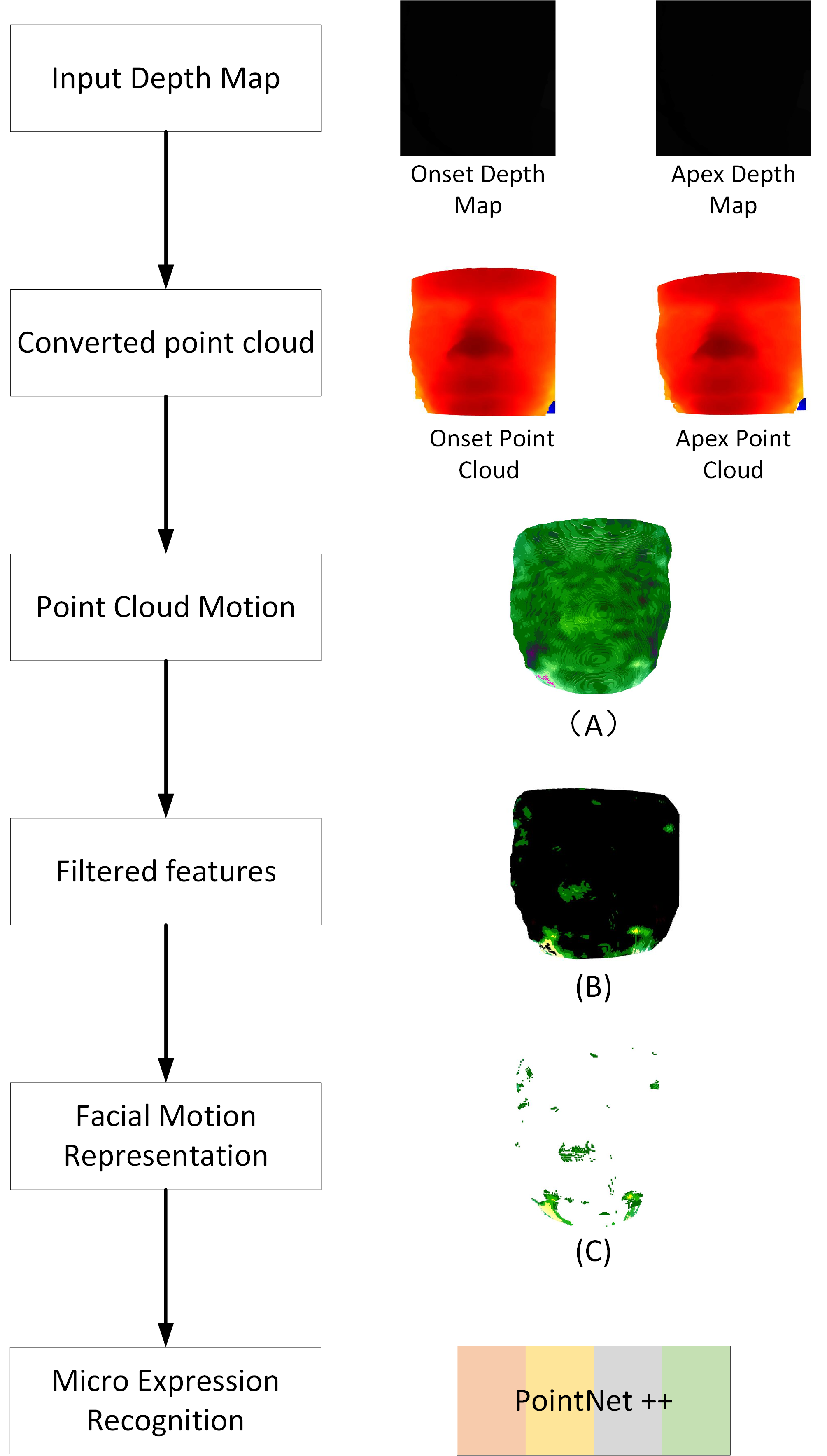}
  \caption{The pipeline of the depth-based MER method entails converting depth maps into point clouds, calculating and preserving the motion features of critical points, and ultimately inputting them into the point cloud network for MER.}
  \label{fig:porcess}
\end{figure}
\subsection{FER with 3D}
3D images provide more reliable geometric information than 2D data. Scaling, rotation, and illumination do not affect extracting certain powerful features from these 3D images \cite{creusot2013machine}.In facial expression recognition, 3D facial meshes and 3D videos (3D mesh sequences, also known as 4D) have been relatively well-researched and summarized\cite{alexandre_systematic_2020,sandbach_static_2012,soltanpour_survey_2017,corneanu_survey_2016,danelakis_survey_2015,li_deep_2022}. Dynamic facial expression recognition can be divided into two main categories: tracking-based and 3D facial model-based. Tracking-based techniques aim to track specific 3D facial model markers using appropriate tracking algorithms. 3D facial model-based techniques aim to exploit facial deformations due to facial expressions. These techniques usually use alignment methods to obtain better results. In the tracking-based techniques, the work of Rosato et al. \cite{rosato2008automatic} dealt with the problem of lack of correspondence of features (or vertices) due to the variable number of vertices in a single model or 3D model sequence for facial expression recognition. A method for automatically establishing vertex correspondence between input scans or dynamic sequences is proposed. Twenty-two feature points are extracted on 2D face textures derived using a deformable template method \cite{yin2001generating}. In \cite{rosato2008automatic}, the composition of descriptors and classifiers is the same as in \cite{chang2005automatic}, but in \cite{rosato2008automatic}, 2D face textures are generated using a corner-preserving mapping and a model adaptation algorithm.\\
And in 3D facial model-based techniques, zhang et al. \cite{yin2001generating} proposed a new 4D spatiotemporal "Nebula" feature to improve the performance of expression and facial motion analysis. The data are voxelized and fitted to a cubic polynomial in a spatiotemporal volume. Labels are specified based on the principal curvature value, and the polarization angle in the direction of minimum curvature is calculated. The labels and angles of each feature are used to construct histograms for each region of the face. The corresponding histogram for each region forms the final feature vector. And a linear discriminant analysis classifier is used to classify.\\
Compared to facial macro-expressions, the motion amplitude of MEs is much weaker, and the 4D FER methods are not applicable. Depth information is a physical representation of facial structure information, and simply feeding the depth map into the network needs to make better use of it. Therefore, we convert the depth map into a point cloud and compute the position change of the point cloud using the onset frame and apex frame on the time series to retain the motion information of the face while filtering the effective points. Compared to 4D FER and image-based MER, our method is more straightforward and more effective in retaining facial motion information and learning expression classification using only the structural information of the face.
\section{Method}
We present a novel method for capturing subtle changes in facial micro-expressions using point cloud representation and propose an effective Micro-Expression Recognition (MER) technique based on it. Our method consists of the following steps: First, we convert depth maps into point clouds to represent facial structures. Second, we compute the point cloud differences between the initial frame and the apex frame as facial motion features. Third, we filter out redundant point clouds to retain effective motion features. Finally, we explore different point cloud models that combine local and global motion features and learn deep hierarchical features to associate expressions with actions. Our method can describe the motion of each pixel in detail, improving robustness, and can be integrated with point cloud classification networks to achieve simple and effective micro-expression recognition. The pipeline is shown in Fig. \ref{fig:porcess}.
\subsection{Preliminaries} Given a ME dataset, $M = {\{(D_i,I_i)\}}_{i=1}^{|M|}$ with $D_i \in \mathbb{R}^{H \times W}$ and $I_i \in \mathbb{R}^{H \times W \times 3}$ where $D_i$ is the depth map of a 2D image $I_i$, and the depth map $D_i$ can be transferred from the pixel location and depth information to the 3D point cloud $P_i \in \mathbb{R}^{N \times 3}$.We aim to capture the motion of the face in the video. For each video sequence $V_i$ in the dataset, we extract $F_o$ = \{($P_o$, $I_o$)\}  and $F_a$ = \{($P_a$, $I_a$)\} of the onset frame $F_o$ and the apex frame $F_a$ to capture facial motion.
\begin{figure*}[h]
  \centering
  \includegraphics[width=\linewidth]{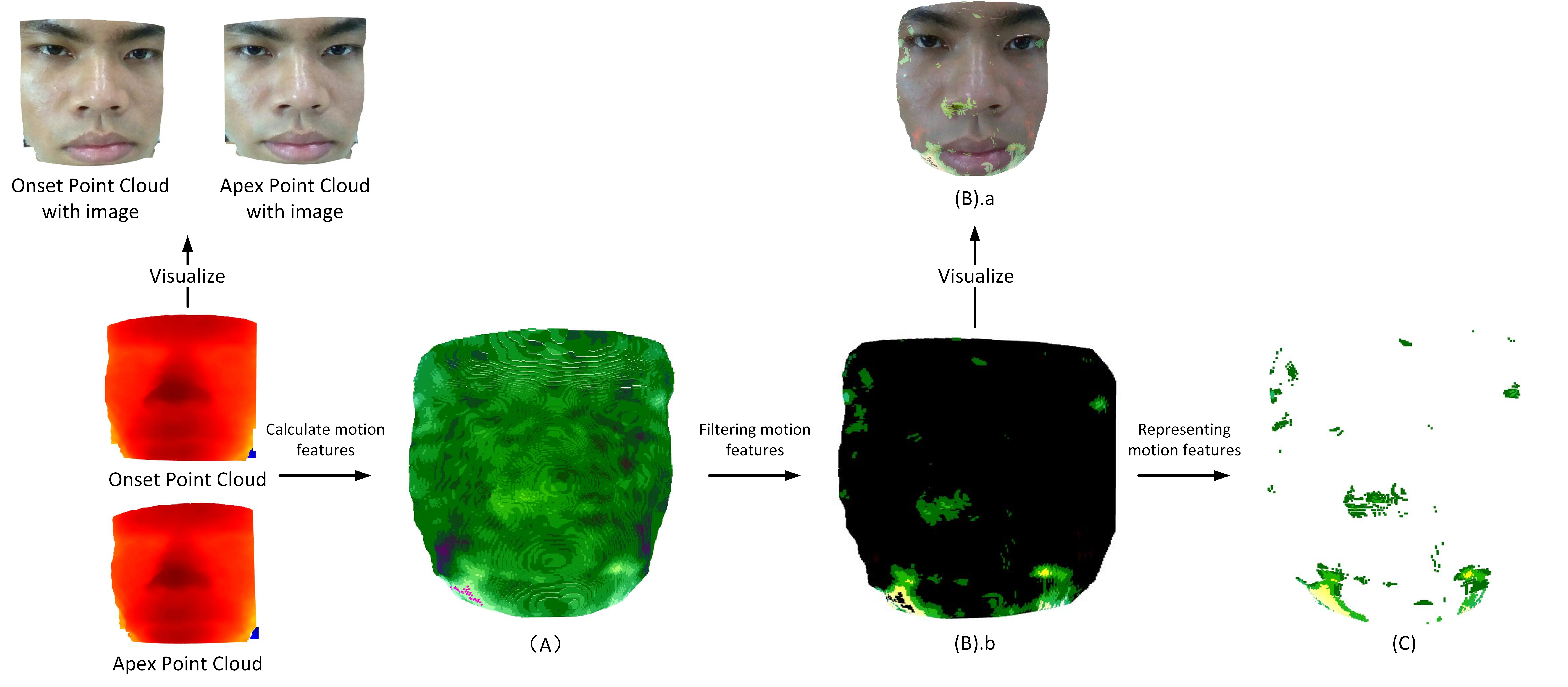}
  \caption{Extraction of facial point cloud motion features. Where (A) is the mapping of the motion of each direction in the color channel, calculated from the point clouds of the onset frame and the apex frame; (B).a is a schematic of the converted motion point cloud mixed with 2D image pixels, which is only used to indicate the motion position; and (B).b is the schematic of the point cloud of the motion after conversion. (C) is the 2048 points of motion extracted after sorting.}
  \label{fig:1}
\end{figure*}
\subsection{Point Cloud Motion Extraction}
The dense point cloud mapped by the depth map preserves the complete facial structure, while the point cloud is highly aligned between frames. To align the faces of the two frames, we first crop and align $F_o$ and $F_a$ using Dlib-ml \cite{king2009dlib}, and then the cropped $D_o$ and $D_a$ are mapped to $P_o$ and $P_a$. Generally, point cloud motion can be computed by scene flow. Still, since the ground motion of MEs is too subtle for scene flow to retain all the motion information accurately, We calculate the difference between two frames of the point cloud, which is the amount of displacement of each point in the three directions $(x, y, z)$. The displacement in these three directions can reflect the intensity and direction of the MEs motion. We used the following equation to calculate the difference:
\begin{equation}
\begin{aligned}
  &\Delta x_i = x_i^a - x_i^o \\ &\Delta y_i = y_i^a - y_i^o \\ &\Delta z_i = z_i^a - z_i^o
\end{aligned}
\end{equation}
Where, $(x_i^o, y_i^o, z_i^o)$ denote the $(x, y, z)$ coordinates of the $i^{th}$ point in $P_o$, $(x_i^a, y_i^a, z_i^a)$ denote the $(x, y, z)$ coordinates of the $i^{th}$ point in $P_a$, and $\Delta P_i = (\Delta x_i, \Delta y_i, \Delta z_i)$ denote the difference of the $i^{th}$ point in the $(x, y, z)$ direction. We combine the $\Delta P = \{P_i\}_{i=1}^N$ with the $P_o$ to represent the ME video sequence $V_i = \{P_o, \Delta P\}$ with $P_o \in \mathbb{R}^{N \times 3}$ and $ \Delta P \in \mathbb{R}^{N \times 3}$. Compared with other feasible methods for calculating point cloud motion, such as scene flow, subtracting two frames of point clouds can preserve motion features simply and efficiently. Every point in the facial structure point cloud obtained from the depth map transformation of two frames has a one-to-one relationship, making it possible to obtain accurate and evident motion features by standardizing the point cloud for subtraction.
\subsection{Motion point cloud filtering}

\begin{figure}[h]
  \centering
  \includegraphics[width=0.7\linewidth]{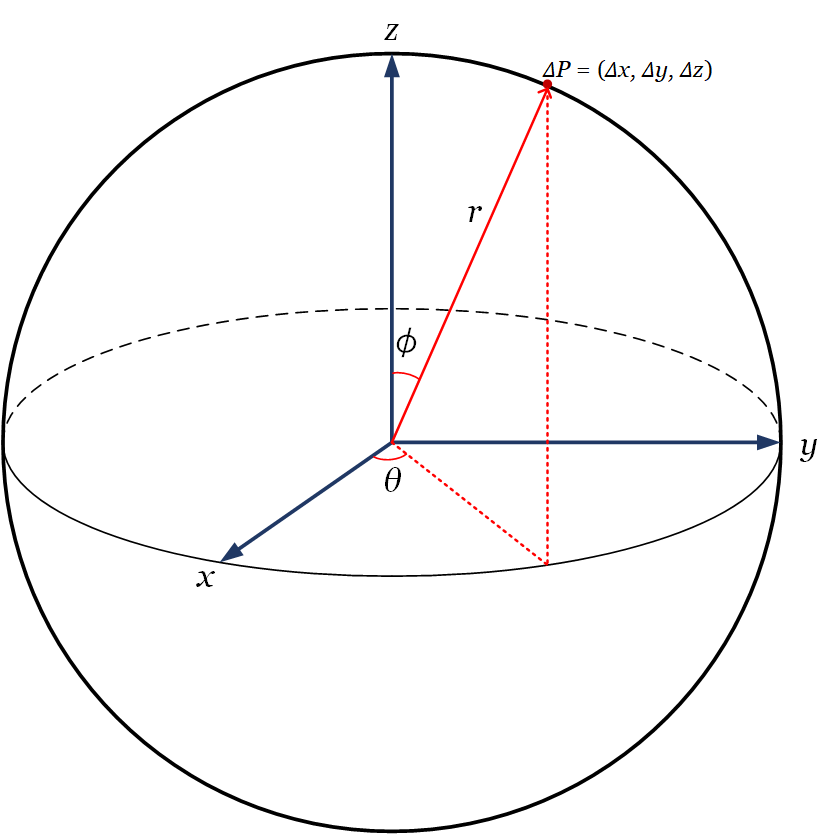}
  \caption{The spatial coordinate to the spherical coordinate diagram. $\theta$ is the angle between the projection of the vector in the $x-y$ plane and $x$, meaning the angle of motion in the $x-y$ plane, and $\phi$ is the angle between the vector and $z$, meaning the angle of motion in the $z$ plane, and $r$ is the mode of three directions,  meaning the amplitude of motion. }
  \label{fig:2}
\end{figure}

Due to micro-expression movements' subtle and covert nature, only a few small regions can produce motion point clouds. In contrast, excessive point clouds without motion information can cause interference. Therefore, we proposed a new method for motion representation to utilize the motion information in the three directions of $\Delta P$, generating a triple channel of color for visualization with Open3D \cite{zhou2018open3d}. This allowed us to represent the motion trends of the face, as demonstrated in Fig. \ref{fig:1} (A). However, two problems arise; firstly, the point cloud of the background often consists of meaninglessly large movements, and secondly, using $\Delta P$ as the color channel to describe facial movements is not intuitive. To counter these problems, we calculated the average distance from each point to the face center, filtering out all distracting points. The filtered point cloud retained the entire face, with distinct motion information at each location. To highlight the motion section of the face, we transformed from spatial Cartesian coordinates to spherical coordinates, as shown in Fig. \ref{fig:2}. This transformation conveyed changes in amplitude and angle of motion more clearly, as depicted in Fig. \ref{fig:1} (B). According to the Facial Action Coding System (FACS), the movement locations and directions of the face collectively represent an expression, so we then calculate the modules $r$ of $\Delta P$ represented the amplitude of motion as the red channel, calculated the angle $\theta$ between the vector and $x$ as the green channel, and the angle $\phi$ between the vector and the $x-y$ plane as the blue channel. The coordinate transformation made facial motion features visible and effectively suppressed the representation of motionless features rendered in black post-normalization, as shown in the changes from Fig. \ref{fig:1} (A) to (B).b, and we calculate $r$ , $\theta$ and $\phi$ in spherical space according to the following equation:
\begin{equation}
\begin{aligned}
  &r = \sqrt{{\Delta x_i} ^2 + {\Delta y_i}^2 + {\Delta z_i}^2} \\ & \theta = \arctan{\frac{\Delta y_i}{\Delta x_i}} \\ & \phi = \arctan{\frac{\Delta z_i}{\sqrt{{\Delta x_i} ^2 + {\Delta y_i}^2}}}
\end{aligned}
\end{equation}
The transformation of coordinates suppresses the representation of motionless features, rendered in black after normalization, and effectively visualizes the motion parts. However, the black point cloud does not contain motion information. Only the point cloud with motion can represent subtle changes in the face. Therefore, we adjusted the number of input point clouds based on the network input size. We calculate the $L2-norm$ of the color channels as the basis for ranking, retaining only the top 2048 points with larger motion amplitudes as input to the network, as shown in Fig. \ref{fig:1}. Finally, we concatenate the retained point cloud positions and colors to represent the motion of ME video sequence $V_i$ where $V_i \in \mathbb{R}^{2048 \times 6}$. The filtered points contain the most significant motion amplitude, direction, and position information, fully representing facial motion features.

\subsection{Point cloud model for MER}
\begin{figure*}[h]
  \centering
  \includegraphics[width=\linewidth]{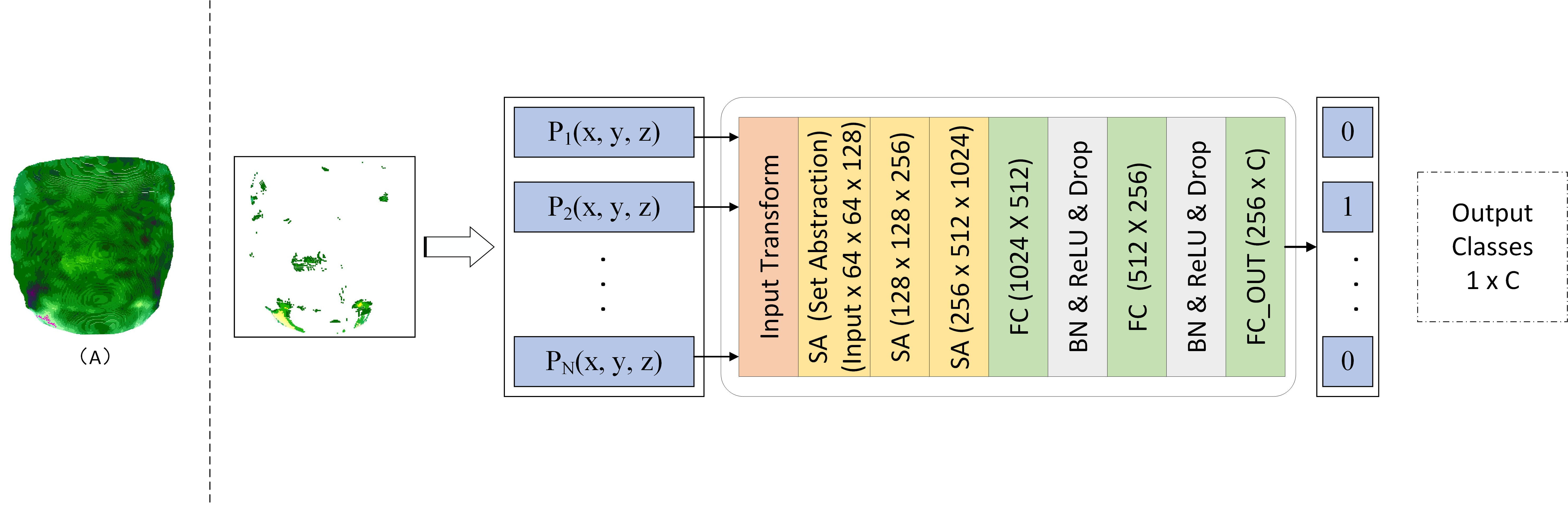}
  \caption{Network structure diagram. The filtered and sorted point clouds are fed into the network, the features are fused and extracted by the Set Abstraction (SA) module, and the classification results are obtained by full connectivity, with C being the number of categories.}
  \label{fig:3}
\end{figure*}

We conducted experiments on a variety of mainstream point cloud models, including networks with convolutional structures, MLP structures, and Transformer structures, to assess the effectiveness of our proposed micro facial expression point cloud representation. We conducted experiments on a variety of mainstream point cloud models, including networks with convolutional structures, MLP structures, and Transformer structures, to assess the effectiveness of our proposed micro facial expression point cloud representation. The results demonstrate the effectiveness of our proposed point cloud representation, with Pointnet++ achieving the best recognition performance.\\
PointNet++ \cite{qi2017pointnet++} is a deep learning method that processes point clouds, which can perform deep hierarchical feature learning on point sets in a metric space. It demonstrates strong processing capabilities in processing point clouds through layer-by-layer aggregation and feature extraction methods. Pointnet-based methods have shown exceptional performance across various tasks \cite{liang2019pointnetgpd}. Concerning ME, PointNet++ applies a hierarchical approach to model local and global aspects of the face, subsequently enhancing sensitivity and robustness to ME.\\
Fig. \ref{fig:3} depicts the network architecture and motion representation. The facial motion is represented by the point cloud consisting of larger motion areas, which serve as input to the network. Studies of facial expressions using the FACS indicate that the different types of facial movement position, amplitude, and direction correspond to different facial expressions reflected in the input of color and position channels. \\
The Set Abstraction (SA) layer performs the task of downsampling, extracting features from the point cloud, dividing it into several subsets, and extracting facial movement position and color channel information for each subset. The feature vectors are then connected to generate overall feature vectors for the point cloud. The SA layer ensures a connection between motion regions and facial expression categories while considering the correlation between motion amplitude and direction. Lastly, the feature vectors are mapped to the ME category space for ME recognition by applying a fully connected layer. Pointnet++ extracts local and global features, models the relationship between motion region and amplitude direction, and improves the robustness of MER.

\section{Experiments}
The inclusion of depth information allows for the detection of structural changes. Since the $CAS(ME)^3$ dataset is the only dataset with depth information available, we evaluated the effectiveness of using depth information to recognize ME. In addition, ablation experiments were conducted on the network parameters to validate the soundness of each part of the design.
\subsection{Dataset}
The dataset comprises multi-modal ME data induced by simulated crime patterns, including depth information, Electrodermal activity (EDA), and sound signals, with a $1280 \times 720$ resolution. During the experiment, depth maps from the $CAS(ME)^3$ dataset of MEs were used to train and evaluate emotion class, providing 1109 labeled MEs and 3490 labeled macro-expressions. \\
The labels for MEs include objective and emotional class, and the labels for macro-expressions include only objective class \cite{davison2018objective}. Objective class is a category defined by facial muscle movement action units (AU) and is categorized into seven groups based on the position of movement combination. Unlike emotion labels, it is characterized by objectivity, consistency, and independence from any person's emotion-based self-reports. \\
On the other hand, emotion labels are analyzed by emotions reported and defined by experts, hence becoming the central aspect of emotional classification. Seven categories represent emotion labels: sadness, anger, disgust, fear, happiness, surprise, and others, with the first four being negative and happiness belonging to the positive.
\subsection{Experimental details}
The dataset samples contain errors and redundancies, and due to occlusion and other reasons, facial expressions of some data were indistinguishable. Thus, we eliminated incorrect samples and preserved 798 annotated ME into four types: positive, negative, surprise, and other. We also extracted three types of 667 ME samples for the experiment, excluding the "other" class.\\
To reduce interference in the background, we used the dlib algorithm to locate the facial region of corresponding depth images from the onset and apex frames of the RGB images. We transformed each pixel's positional and depth information into two frames of point clouds, the depth camera focal length is 1324.65, and the scaling factor is 1000; we retain motion information in all directions while filtering out point clouds beyond 1.5 times the average center-point length. This process not only retained the facial structure but also reduced background interference. For the color channels, we normalize each of the three channels while filtering out distorted motion features to prevent the effects caused by the offset of the overall face. Lastly, we chose 2048 points with the most prominent motion features. \\
For PointNet++, Adam is used as the optimizer, with a weight decay of 0.0001, batch size set to 24, and learning rate initialized to 0.001. The input size is 2048 point clouds, including both position and color channels, containing motion positions, amplitudes, and directions. 
To ensure the accuracy of the test,  the traditional leave-one-subject-out approach (LOSO) was used in the testing phase. The unweighted F1 score ($UF1$) and the unweighted average recall ($UAR$) were used as performance metrics to avoid over-fitting the proposed method to a particular class. The number of true positives ($TP_{C}$,  false positives ($FP_{C}$), and false negatives ($FN_{C}$) for each class $c$ (where $C$ is the total number of classes) is assumed to exceed the number of subjects,  and $UF1$ is calculated as 
	\begin{equation}
		UF1 = \sum\nolimits_i^C {UF{1_i}/C}
		\label{eq:9}
	\end{equation}
	where
	\begin{equation}
		UF{1_c} = \frac{{2*T{P_c}}}{{2*T{P_c} + F{P_c} + F{N_c}}}
		\label{eq:10}
	\end{equation}
	and $UAR$ can be expressed as
	\begin{equation}
		UAR = \frac{1}{C}\sum\nolimits_i^C {Ac{c_c}}
		\label{eq:11}
	\end{equation}
	with
	\begin{equation}
		Ac{c_c} = \frac{{T{P_c}}}{{{n_c}}}
		\label{eq:12}
	\end{equation}
\subsection{Experimental results and analysis}
To ascertain the efficacy of utilizing depth information for the micro-expression recognition, we comparative analysis was conducted against state-of-the-art methods which . The results demonstrate significant improvements, , as shown in Table \ref{tab:1} of $CAS(ME)^3$. 
In order to fairly validate the effects of each emotion category, we tested using emotions of the 3, 4, and 7 emotion classes, and our method outperformed the compared methods by 10 \% - 20 \%  in each category (bolded in Table \ref{tab:1}). In the three emotion classes, our method exhibits a more significant improvement, achieved 61.42\% $UF1$ and 67.10\% $UAR$, which is more than 20\% better than the RCN-A \cite{xia2020revealing} method; for the four emotion classes, our method achieved 43.32\% $UF1$ and 49.24\% $UAR$, which is about 10\% better than the Baseline(+Depth) \cite{li2022cas}; for the seven emotion classes, our method also outperformed the Baseline(+Depth) \cite{li2022cas} method by about 10\%, achieving 30.59\% of $UF1$ and 32.43\% $UAR$.\\ In the four and seven emotion classes, the improvement is comparatively smaller. Possible reasons include insufficient data and an imbalanced sample, leading to the model's incapacity for effectively addressing the challenges associated with multiple classifications. Furthermore, the high similarity in feature representation between different emotional classes of micro-expressions obstructs straightforward differentiation, necessitating a larger dataset and a more nuanced feature representation.
\begin{table}[h]
\caption{MER on the $CAS(ME)^3$ dataset.}
\label{tab:1}
\centering
\begin{tabular}{lccc}
\hline
Method           & Classes    & $UF1$ (\%)       & $UAR$ (\%)       \\ \hline
FR \cite{zhou2022feature}               & 3          & 34.93          & 34.13          \\
STSTNet \cite{liong2019shallow}          & 3          & 37.95          & 37.92          \\
RCN-A \cite{xia2020revealing}           & 3          & 39.28          & 38.93          \\
\textbf{Ours}    & \textbf{3} & \textbf{61.42} & \textbf{67.10} \\ \hline
Baseline \cite{li2022cas}        & 4          & 29.15          & 29.10          \\
Baseline(+Depth) \cite{li2022cas} & 4          & 30.01          & 29.82          \\
\textbf{Ours}    & \textbf{4} & \textbf{43.32} & \textbf{49.24} \\ \hline
Baseline \cite{li2022cas}         & 7          & 17.59          & 18.01          \\
Baseline(+Depth) \cite{li2022cas} & 7          & 17.73          & 18.29          \\
\textbf{Ours}    & \textbf{7} & \textbf{30.59} & \textbf{32.43} \\ \hline
\end{tabular}
\end{table}
\begin{figure*}[h]
  \centering
  \includegraphics[width=\linewidth]{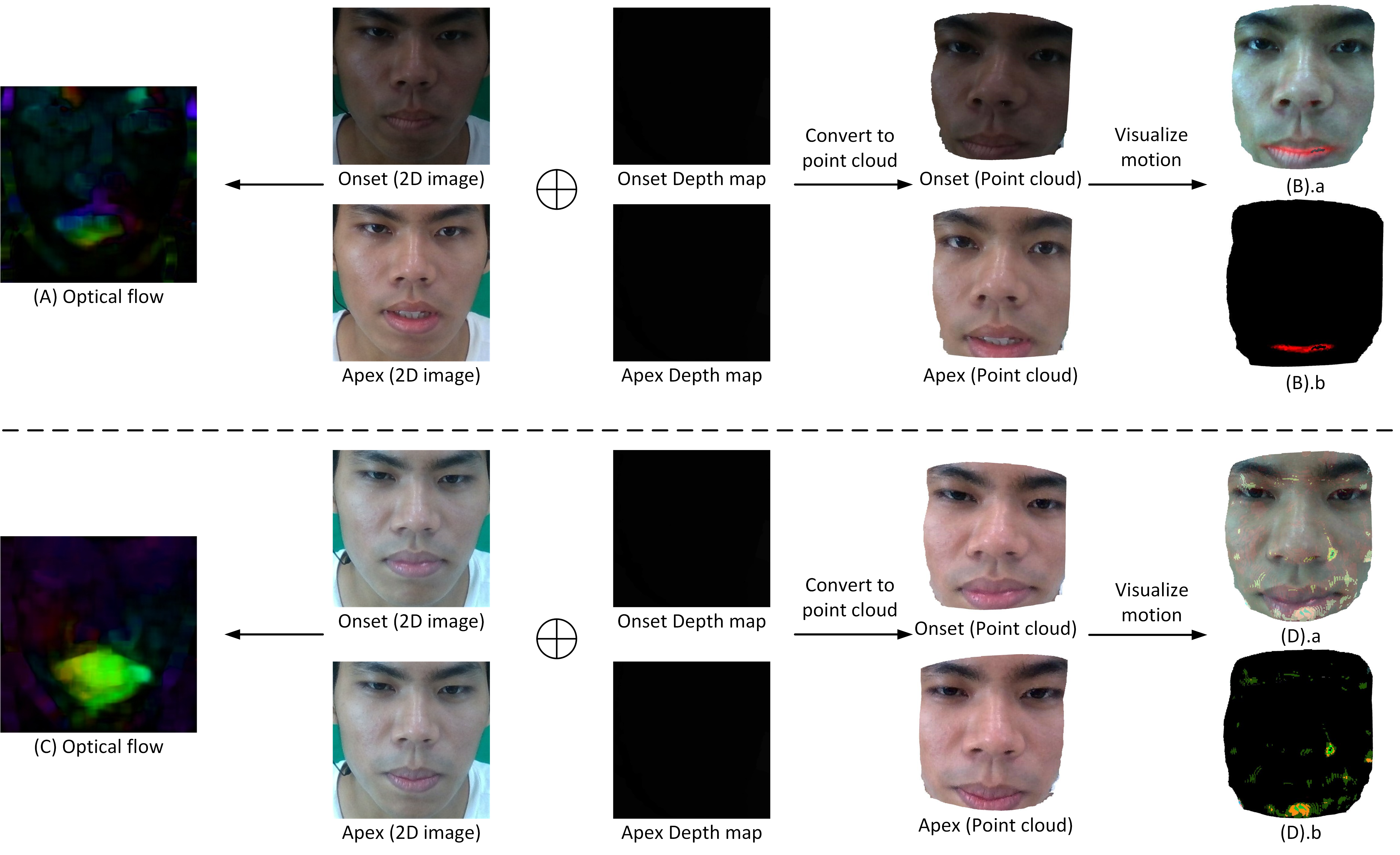}
  \caption{Comparison of optical flow features and point cloud motion features. Where (A) and (C) are the optical flow maps of the extracted 2D image onset and apex frames. The point cloud is obtained by converting the depth information corresponding to the 2D image. (B) and (D) are the extracted point cloud motion features, where (B).a and (D).a used only to illustrate motion position; and (B).b and (D).b for further extraction of motion position.}
  \label{fig:4}
\end{figure*}\\
We conducted experiments on multiple point cloud networks, namely KP-Conv \cite{thomas2019kpconv}, PointNet \cite{qi2017pointnet}, Point Transformer (PT-v1) \cite{zhao2021point}, and Pointnet++. These experiments aimed to validate the effectiveness of the proposed micro-expression point cloud representation. Interestingly, all of the tested networks outperformed both the baseline and the current state-of-the-art methods. The comparative results are presented in Table \ref{tab:2}. And the first row in Table \ref{tab:2} shows the optimal results achieved using the image.
\begin{table}[h]
\caption{Multiple model validation of micro-expression point cloud representation effectiveness.}
\label{tab:2}
\centering
\begin{tabular}{cccc}
\hline
Method              & Backbone           & UF1(\%)        & UAR(\%)        \\ \hline
RCN-A   \cite{xia2020revealing}            & Image-RCN          & 39.28          & 38.93          \\
PointNet \cite{qi2017pointnet}           & Point-MLP          & 58.58          & 64.27          \\
KP-Conv \cite{thomas2019kpconv}            & Point-CNN          & 59.51          & 66.23          \\
PT-v1 \cite{zhao2021point}  & Point-Transformer  & 55.22          & 63.42          \\
\textbf{PointNet++} \cite{qi2017pointnet++} & \textbf{Point-MLP} & \textbf{61.42} & \textbf{67.10} \\ \hline
\end{tabular}
\end{table}\\
 Our proposed micro-expression point cloud representation method has outperformed the current state-of-the-art methods in multiple exemplary point cloud models. To compare the disparities between local and global modeling, we conducted experiments using PointNet, a local modeling method, and PointNet++, a global modeling method. PointNet achieved $UF1$ of 58.58\% and $UAR$ of 64.27\%, whereas PointNet++ achieved $UF1$ of 61.42\% and $UAR$ of 67.10\%. \\
 The findings suggest that PointNet++, with a global modeling approach, can more effectively establish associations between various motion positions and expressions, thereby capturing the relationships between motion positions and expressions and achieving optimal micro-expression recognition results (bold in Table \ref{tab:2}). \\
 Additionally, we performed experiments on KP-Conv, a technique based on point convolution, and PT-v1, a method using Transformers. KP-Conv obtained $UF1$ of 59.51\% and $UAR$ of 66.23\%, whereas PT-v1 achieved $UF1$ of 55.22\% and $UAR$ scores of 63.42\%. These results closely resemble those obtained by PointNet, below PointNet++. It is possible that the presence of certain parameters, like the sampling rate, and the relatively sparse nature of micro-expression point clouds, necessitate meticulous adjustment of network parameters. \\
 Moreover, it is necessary to scale the network based on the sample size. Nevertheless, the results reaffirm the ability of our proposed micro-expression point cloud representation method to effectively retain the motion and semantic information of micro-expressions and accurately discern micro-expressions.
\par{\textbf{So why can depth map work well ?}} \\Initially, the preprocessing of the depth map is crucial. The changes in facial depth correlate to changes in facial structures. The spontaneous movement of MEs is quick and brief local muscle movements \cite{dong2022intentional}. The relative variations in facial structures are more distinct in this case. Facial cropping and normalization of motions are required to address interference from the background and shift in the face. By transforming the point motion coordinate system, the method can preserve the vector characteristics of motion direction and magnitude while filtering out non-motion points, thereby retaining the full motion information at crucial positions.\\
Depth maps offer unique advantages over 2D images. Changes in lighting conditions do not impact depth maps and are highly proficient in detecting motion positions. Furthermore, depth maps can capture motion features, which are challenging to identify through texture information, resulting from the motion of facial structures. \\
However, images or optical flow inputs are vulnerable to lighting and facial displacement changes. Despite the ability of the optical flow to retain motion features, variations in lighting can result in regions of the face lacking motion features in the acquired optical flow, as shown in Fig. \ref{fig:4} (A), the optical flow map of the onset and apex 2D frames reveals that due to changes in lighting, the onset frame appears darker. There are two problems, one is the weak optical flow feature of the mouth's correct motion position (labeled as AU25, with separation of lips and exposure of teeth) in the optical flow map, and the other is significant background interference. In contrast, sensors that capture depth information are insensitive to lighting conditions, and movement position descriptions through 3D point clouds are more accurate and robust than those from 2D images (Fig. \ref{fig:4} (B)), the motion position obtained is more concentrated and obvious.\\
Furthermore, the optical flow encounters difficulty in capturing these minute movements due to unclear alterations in some facial area pixel values(Fig. \ref{fig:4} (C)). The optical flow map extracts only the motion feature of the whole mouth (labeled AU16+24, lower lip pulled down, and lips squeezed together), and other positions of insignificant motion are not concerned. Conversely, depth maps can competently capture delicate variations in facial structures provoked by muscular motions, as shown in Fig. \ref{fig:4} (D). The depth map can accurately capture the movement position of the lower lip and discover the subtle movement on the right side of the nose that optical flow cannot detect, providing a new method for observing the correlation between movement areas and emotions.
\subsection{Ablation studies}
The selection of input points is a crucial aspect of the experiment. Points with greater motion amplitude contain more semantic information, rendering this selection crucial. In addition, the number of input points must also be considered. For this reason, we conducted ablation experiments to examine the effects of input point selection and numbers, as shown in Table \ref{tab:3}.\\
\begin{table}[h]
\caption{The selection and number of points.}
\label{tab:3}
\centering
\begin{tabular}{ccccccc}
\hline
\multicolumn{2}{c}{Point selection}                            & \multicolumn{3}{c}{Point number}                                                        & \multicolumn{2}{c}{Evaluation Metrics} \\ 
Sort                               & Random                    & 1024                      & 2048                               & 4096                      & $UF1$(\%)          & $UAR$(\%)         \\ \hline
\textbf{\checkmark} & \textbf{}                 & \textbf{}                 & \textbf{\checkmark} & \textbf{}                 & \textbf{61.42}     & \textbf{67.10}    \\
                                   & \checkmark &                           & \checkmark          &                           & 43.69              & 47.51             \\ \hline
\checkmark          & \textbf{}                 & \checkmark & \textbf{}                          &                           & 54.70              & 57.43             \\
\textbf{\checkmark} & \textbf{}                 & \textbf{}                 & \textbf{\checkmark} & \textbf{}                 & \textbf{61.42}     & \textbf{67.10}    \\
\checkmark          &                           &                           &                                    & \checkmark & 56.25              & 58.49             \\ \hline
\end{tabular}
\end{table}

\begin{figure}[h]
  \centering
  \includegraphics[width=\linewidth]{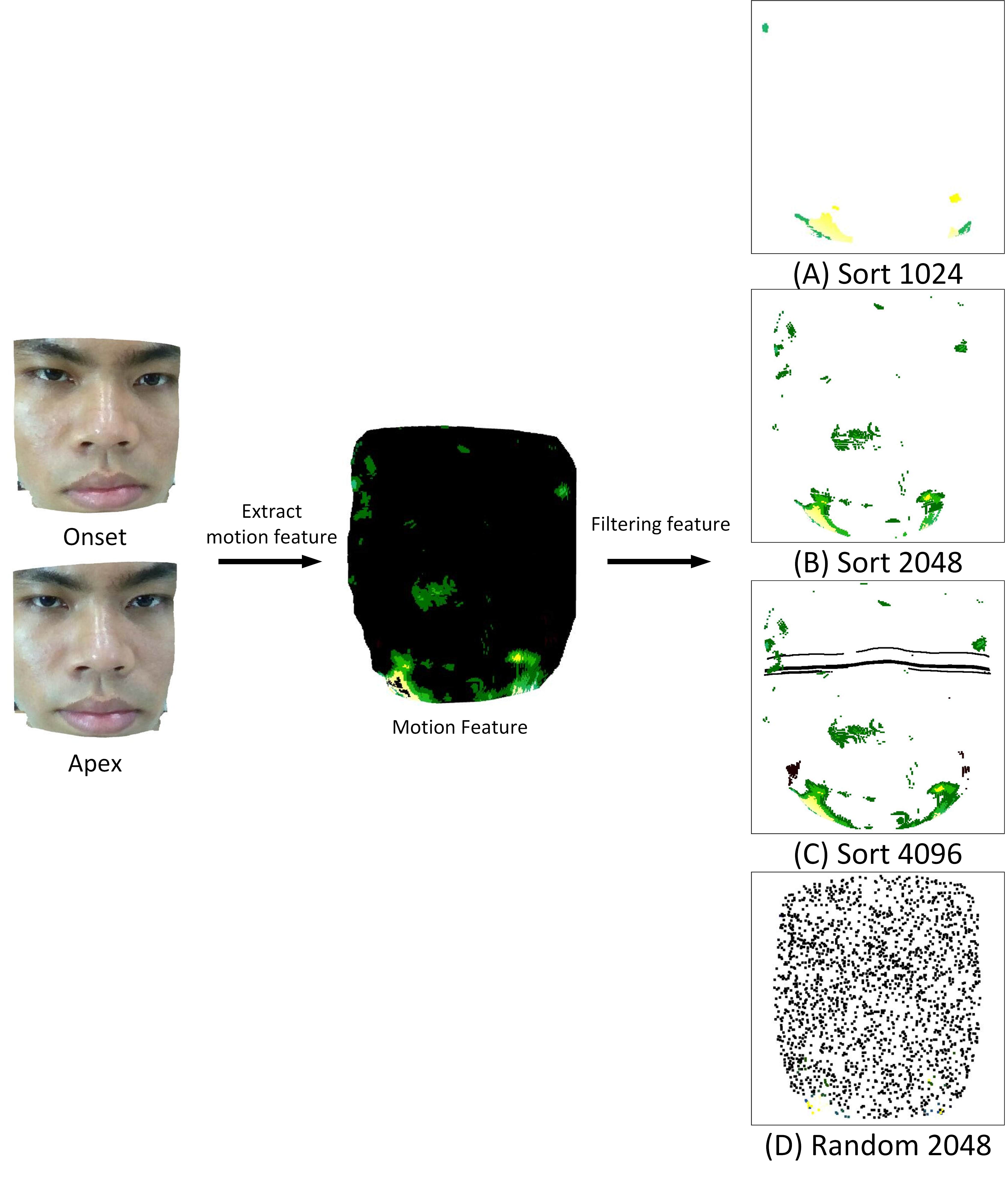}
  \caption{Motion point number and selection method. Where (A) is sorted to extract 1024 points, (B) is sorted to extract 2048 points, (C) is sorted to extract 4096 points, and (D) is random to extract 2048 points.}
  \label{fig:5}
\end{figure}

We evaluated the selection methods and quantity of points under three emotion categories. The most optimal result was achieved by selecting and sorting 2048 points (bolded in Table \ref{tab:2}). When using only random sampling, the $UF1$ and $UAR$ were only 43.69\% and 47.51\%, respectively, indicating reduced recognition accuracy. Since the random sampling points were distributed consistently, it wasn't easy to maintain facial structural and motion position information.Fig. \ref{fig:5}. (D) demonstrated that inputs generated through random sampling have less semantic information overall, confirming the validity of sorting the motion information as an input.\\
Furthermore, we also assessed the results of different quantities of sorted points. Due to the minimal quantity, 1024 points resulted in $UF1$ and $UAR$ of 54.70\% and 57.43\%, respectively, and failed to retain complete motion and position information, as shown in Fig. \ref{fig:5}. (A).; with 4096 points, $UF1$ and $UAR$ of 56.25\% and 58.49\%, respectively, were derived, resulting in a surplus of data with too much position information and without motion semantics, as shown in Fig. \ref{fig:5}. (C). After sorting, $UF1$ and $UAR$ of 61.42\% and 67.10\%, respectively, were obtained with 2048 points, achieving the optimal recognition effect by holding an appropriate amount of motion information, as demonstrated in Fig. \ref{fig:5}. (B).
\subsection{Objective class evaluation}
In addition to using emotional class as classification criteria, we attempted to evaluate based on objective class. Objective class grouped facial movements based solely on the movement position, immune to subjectivity from the subjects and experts. Before input, we extracted the full facial motion positions. The network could efficiently complete the objective class's seven-class classification by using our method, as shown in Table \ref{tab:4}.\\
\begin{table}[h]
\caption{Objective class evaluation.}
\label{tab:4}
\centering
\begin{tabular}{cccc}
\hline
\multicolumn{2}{c}{Training method} & \multicolumn{2}{c}{Evaluation Metrics} \\
Train            & Test             & $UF1$(\%)            & $UAR$(\%)           \\ \hline
Micro            & Micro            & 51.37              & 61.42             \\
Macro            & Micro            & 63.28              & 61.55             \\ \hline
\end{tabular}
\end{table}
We evaluated \textbf{seven-class} for the objective class using the leave-one-subject-out method to train and evaluate ME and achieved $UF1$ 51.37\% and $UAR$ 61.4\%; Significant problems with over-fitting and sample distribution imbalance were observed in the network. However,  by training the macro-expression samples and testing on the ME samples without using any transfer strategies or the leave-one-out method and ME samples were not trained, we managed to attain $UF1$ of 63.28\% and $UAR$ of 61.55\%, which significantly alleviated problems associated with over-fitting and sample distribution. Because the objective class is suited well for our method, and the transfer learning method remediated the problem of insufficient sample, we can obtain over 60 \% accuracy even in seven classifications.
\section{Limitation}
The limitations of our method are twofold: limitations of MER tasks and limitations using depth maps. Before the $CAS(ME)^3$ dataset, there were two main problems with MER tasks using 2D images: the need for datasets and the uneven distribution of samples. While the $CAS(ME)^3$ dataset offers many 2D and depth map samples for micro and macro-expressions, macro-expressions lack emotional label annotations, and other data sets do not contain depth map samples. For methods that rely solely on depth maps, however, there still exists a scarcity of depth map samples; this scarcity can give rise to over-fitting and needs to be readily addressable through transfer learning methods. Results from objective class evaluation experiments have demonstrated that the problems of over-fitting and uneven sample distribution can be mitigated through transfer learning methods using macro-expression samples.\\
Although our approach has surpassed the baseline method, much room remains for improvement:
The depth map only reflects the movement of facial structures, while the changes in facial textures and pixel regions can also provide rich semantic information.
The strategies currently used in the method need a solid theoretical foundation. Hopefully, they can be improved in the future.
The overall facial structure also contains certain semantic information, but due to the limitations of network input and data redundancy, it is impossible to input complete facial structure information.
Using complete facial structure information is possible while preserving the motion information.

\section{Conclusion AND FUTURE WORK}
We propose a MER method that relies exclusively on depth maps. The method focuses on the motion information of the facial position and the direction of facial motion amplitude to establish a relationship between the motion position and expression class. Transforming and sampling point clouds and using point cloud models with superior point cloud processing capabilities to recognize micro-expressions.The experiment revealed that our method could accurately and robustly locate motion positions; increasing the number of samples leads to better recognition results. These promising results suggest that better results can be attained by increasing the quantity of depth or point cloud data.\\
We intend to introduce additional methods to address the issue of insufficient samples. Semi-supervised and self-supervised methods are among the methods that are expected to provide a solution. Transfer learning is also an effective method. Additionally, we plan to improve the recognition performance of the network by combining features from the pixel region, texture variations, and entire facial structures.
\section*{Acknowledgments}
This work was supported partly by the Natural Science Foundation of Hainan Province (Grant No. 622RC675), National Natural Science Foundation of China (Grant No. 62173045, 62273054), partly by the Fundamental Research Funds for the Central Universities (Grant No. 2020XD-A04-3), and the BUPT Excellent Ph.D. Students Foundation (Grant No. CX2023115).
\bibliographystyle{IEEEtran}
\bibliography{samples/acmart}

\end{document}